\documentclass[amsmath, amsfonts, amssymb,twocolumn]{revtex4}

\usepackage{graphicx}
\usepackage[usenames]{color}
\usepackage{ulem}

\begin{document}

\title{Self-localized states in species competition}

\author{Pavel V. Paulau$^{1,2}$, Dami\`a
Gomila$^1$\footnote{Email address: damia@ifisc.uib-csic.es;
telephone: +34 971259837; fax: +34 971173248.}, Crist\'obal
L\'opez$^1$ and Emilio Hern\'andez-Garc\'{\i}a$^1$}

\affiliation{$^1$IFISC (CSIC-UIB), Instituto de F\'isica Interdisciplinar y
Sistemas Complejos, Campus Universitat de les Illes Balears,
E-07122 Palma de Mallorca, Spain \\
$^2$ Institute for Chemistry and Biology of the Marine Environment (ICBM), 
Carl-von-Ossietzky University of Oldenburg, Carl-von-Ossietzky-Strasse 9-11, 
26111 Oldenburg, Germany
}

\date{January 21, 2014}

\begin{abstract}
We study the conditions under which species interaction, as
described by continuous versions of the competitive
Lotka-Volterra model (namely the nonlocal Kolmogorov-Fisher
model, and its differential approximation), can support the existence of
localized states, i.e. patches of species with enhanced
population surrounded in niche space by species at smaller
densities. These states would arise from species interaction,
and not by any preferred niche location or better fitness. In
contrast to previous works we include only quadratic
nonlinearities, so that the localized patches appear on a
background of homogeneously distributed species coexistence,
instead than on top of the no-species empty state. For the
differential model we find and describe in detail the stable localized
states. For the full nonlocal model, however competitive
interactions alone do not allow the conditions for the
observation of self-localized states, and we show how the
inclusion of additional facilitative interactions lead to the
appearance of them.
\end{abstract}

\maketitle

\section{Introduction}
\label{Introduction}

The interactions among the biological entities integrating an
ecosystem give rise to surprising emergent behaviors.
Competition is one of the most important and ubiquitous of
these interactions: if there is an increase in the population
of one species, due to the consumption of common resources or
shared predators \cite{Holt1977}, there is a decrease in the
growth rate of the others. Because of this interaction it is
usually argued that a given ecosystem can host only a limited
number of species that should be sufficiently separated from
each other in the so-called niche space. This is the
$d$-dimensional space whose coordinates ($x$, $y$, \dots)
quantify the traits of the species relevant for the utilization
of the resources distributed as a function of these
coordinates. The {\it competitive exclusion principle}
\cite{Abrams1983} is a formulation of this situation, in which
species can not coexist too close in niche space ({\it limiting
similarity}). Despite this reasoning, however, it should be
said that even the most traditional mathematical model of
competitive species, the Lotka-Volterra (LV) model
\cite{Murray1993}, is known to allow solutions characterized by
a continuous distribution of species \cite{Roughgarden1979} under some 
circumstances (see reviews in \cite{Leimar2013,Barabas}). More remarkable in 
this context are recent results on the LV model (or closely related
ones) showing the existence of solutions that do not represent
purely continuous coexistence, nor are typical of a limiting
similarity situation \cite{Scheffer2006, Pigolotti2007,
HernandezGarcia2009, Leimar2013}: clusters of species around
particular niche positions well separated from each other and
filling out the niche space. For the competitive LV model,
these lumped distributions appear due to pattern forming
instabilities triggered by the shape of the interaction
function \cite{Pigolotti2007, HernandezGarcia2009,
Fort2009,Pigolotti2010}. Besides these many-cluster species
configurations, an ecologically relevant question is under
which conditions solitary clusters of species may appear. These
would arise from an evolutionary or random drift towards a
particular niche position, or simply from an advantageous
initial condition. In this Paper, we address this question in
the context of pattern formation in continuous versions of the
LV model, both in an integral formulation as in its
differential approximation. Our focus is on competitive
interactions, but we will be forced to consider also some
facilitative (i.e. mutualistic or symbiotic) situations.
Through the paper we will keep in mind the situation of species
competition in niche space, but we stress that the concepts and
type of models used here are equally valid to describe
organisms randomly moving in physical space and nonlocally
competing for resources with other individuals in their spatial
neighborhood \cite{HernandezGarcia2004,Lopez2004}, or rather in
evolutionary situations \cite{Doebeli2007, Leimar2008}.

A pattern-forming instability or bifurcation is a source of
great complexity, and many different scenarios may arise from
it. One of the simplest cases is the formation of a periodic
structure, that in two or higher dimensions can have different
geometries depending on the nature of the nonlinearities. In
some cases the bifurcation can be subcritical so that periodic
patterns can coexist with homogeneous distributions. In this
case localized solutions consisting of one or more isolated
lumps on top of a homogeneous distribution might exist, being
supported by the nonlinearity and the spatial coupling, as
shown in general amplitude equations
\cite{Woods1999,Coullet2000}. If this mechanism turns out to be
present in the context of biological competition, then a stable
localized lump could be formed in a given stable ecosystem
supporting a continuous coexistence of species. Such lumps can
be formed at any position in niche space triggered by
particular perturbations or initial conditions. This means that
species with no special advantage with respect to their
competitors might prevail at some point due to a particular
initial condition. These high values of the population of
certain species would be supported by the nonlinear dynamics
and the spatial interaction, and not by a better fitness to the
ecosystem.

The LV competition model in niche space turns out to be a nonlocal
model, i.e., species interact with others not located closely
in the niche axis. Population dynamics has revealed many
different interesting phenomena due to nonlocal competition
\cite{Fuentes2003,Maruvka2006,Baptestini2009,Pigolotti2007,
Clerc2005,Clerc2010}, such as periodic patterns, discrete
clusters, defects and fronts in space, etc. Self-localization has been
broadly studied in physical systems \cite{Descalzi2011,Akhmediev2008,Jacobo}
but much less in the context of population dynamics
\cite{Escaff2009,Clerc2005,Clerc2010}.

Previous works have already found self-localization of
biological entities by inclusion of the Allee effect, i.e. a
tendency to extinction when population numbers are too small,
in nonlocal competition models
\cite{Escaff2009,Clerc2005,Clerc2010} with cubic nonlinearity.
The Allee effect naturally induces bistability between the
empty or extinct state and the natural occupation determined by
the carrying capacity. This bistability allows the existence of
self-localized patches of densities close to the carrying
capacity surrounded by empty space. In this case the
bistability involves two different spatially homogeneous states
\cite{Burke2006}. In this Paper, in contrast, we address the
situation involving coexistence of a spatially homogeneous
state and a spatially periodic pattern
\cite{Woods1999,Coullet2000}. Also, we consider always positive
linear growth rates, so that the Allee effect is absent and a
small population will always grow, and we use only quadratic
nonlinearities. In consequence, we are looking for localized
structures on top of a non-zero homogeneous density, instead of
the localization on top of an unpopulated background as
described previously \cite{Escaff2009,Clerc2005,Clerc2010}.
Thus, we are considering the possibility that the interaction
enhances the density locally, but without driving to extinction
the rest of the system.

The structure of this article is as follows: In Section
\ref{model} our nonlocal model for species competition is
introduced. In section \ref{truncation} we approximate this
model by a partial differential equation (PDE) which reproduces
the basic original results, and allows us to use methods for
the analysis of self-localized solutions in PDEs. In Section
\ref{results} we present the results of our analysis for this
differential model showing under which conditions localized
solutions can be found. Then, in Section \ref{backtononlocal}
we discuss the features the nonlocal interaction kernel must
have to observe localized states in the full nonlocal model.
Finally, in Section \ref{conclusions} we give some concluding
remarks. The Appendix briefly summarizes details of the
numerical methods.

\section{The nonlocal Kolmogorov-Fisher model for species competition}
\label{model}

The classical Lotka-Volterra model of $N$  species in
competition, each utilizing a common distributed resource $x$
is given by \cite{Roughgarden1979,Pigolotti2007,Scheffer2006}
\begin{equation}
\label{LogisticEquation}
\dot{n_i}=n_i \left( r - \sum_{j=1}^N G(x_i-x_j) n_j\right) \ i=1,..., N,
\end{equation}
where the dot denotes temporal derivative, $n_i$ is the
population of species $i$, $r$ is the growth rate (that we
assume the same for all species), and $x_i$ is the position of
the species $i$ in the niche axis (for simplicity we work in
one dimension). $G(x)$ is the interaction kernel, which unless
explicitly said will take positive values to model competitive
interaction. We also assume $G$ to depend only on the modulus
of the relative difference $|x_i-x_j|$, meaning that resources
are homogeneously distributed in niche space and interactions
are isotropic there. $G$ sets the scale of the carrying
capacity, which is then also the same for all niche positions.
More complex situations are reviewed in \cite{Leimar2013}.

If niche locations are considered to form a continuum (the
infinite real line), we can write the former equation as:
\begin{equation}
\dot{\Psi} = \Psi (r - \tilde{G} \Psi), \\
\label{LotkaVolterraContinuous}
\end{equation}
where $\Psi(x)$ is now the population density (always
positive), and $\tilde{G}$ is an integral operator describing
the  competition term:
\begin{equation}
\tilde{G} \Psi = \int\limits_{-\infty}^{+\infty} G(x-s) \Psi(s) ds. \\
\label{Goperator}
\end{equation}

A further step in the modeling is considering diffusion in
niche space, that may account, for instance, for mutations in an
evolutionary context,
or random phenotypic changes
\cite{Lawson2007bmb,Pigolotti2010}:
\begin{equation}
\dot{\Psi}(x) = \Psi(x) ( r - \tilde{G} \Psi ) + D \frac{\partial^2 \Psi(x)}{\partial x^2}, \\
\label{KolmogorovFisherModel}
\end{equation}
where $D$ is the diffusion coefficient. Note that Eq.
(\ref{KolmogorovFisherModel}) is a type of nonlocal
Kolmogorov-Fisher-like equation
\cite{Britton1989,Sasaki1997,Fuentes2003,HernandezGarcia2004,Genieys2006,Perthame2007}.
This type of equation may also describe organisms randomly
moving in physical space and nonlocally competing with other
individuals for resources \cite{HernandezGarcia2004,Lopez2004}.
In that case $D$ is a true diffusion coefficient modeling
random dispersion in space.

It has been shown that arbitrarily small 
structural perturbations of this model
away from having a constant $r$ may destroy the continuous all-positive
solution for zero diffusion \cite{Barabas,Barabas2}. The presence of diffusion 
in our case ensures, however, that the homogeneous solution only deforms 
continuously under small perturbations from a constant $r$. In this case the 
existence and dynamical properties of self-localized states are not drastically 
altered, as shown, for instance, in a nonlinear optical system \cite{Jacobo}.

\section{Truncation of the nonlocal operator}
\label{truncation}

In order to analyze the existence of localized states in Eq.
(\ref{KolmogorovFisherModel}), we first approximate the
nonlocal operator (\ref{Goperator}) by a simpler differential
operator. This will allow us to apply standard techniques for
PDEs to find localized states. To do so we Taylor expand the
function $G$ in the nonlocal operator to obtain a series of
derivatives of $\Psi$:
\begin{equation}
\tilde{G} \Psi = G_0 \Psi + G_1 \frac{\partial \Psi}{\partial x} + G_2 \frac{\partial^2 \Psi}{\partial x^2} +
 G_3 \frac{\partial^3 \Psi}{\partial x^3} + G_4 \frac{\partial^4 \Psi}{\partial x^4} + ... , \\
\label{GoperatorDecomposed}
\end{equation}
where
\begin{equation}
G_n = \frac{(-1)^n}{n!} \int\limits_{-\infty}^{+\infty} G(z) z^n dz.
\label{Gcoefficients}
\end{equation}
Because of the assumed isotropy of $G$, $G=G(|z|)$, all terms
$G_n$ with odd values of $n$ are zero.

The $k$-Fourier component of the convolution integral operator
can be written as:
\begin{equation}
\widehat{\left \{ \tilde{G} \Psi \right \}}_k = \hat{G}(k) \hat{\Psi}(k), \
\label{GoperatorFourier}
\end{equation}
where the hat indicates Fourier transform. From Eq.~
(\ref{GoperatorDecomposed}), one can also find (for isotropic
systems) that:
\begin{equation}
\widehat{\left \{ \tilde{G} \Psi \right \}}_k = G_0 \hat{\Psi} (k) -
G_2 k^2 \hat{\Psi}(k)  + G_4 k^4 \hat{\Psi}(k) + ...,  \\
\label{GoperatorDecomposedFourier}
\end{equation}
therefore
\begin{equation}
\hat{G}(k)= G_0  - G_2 k^2  + G_4 k^4 + ...  \\
\label{GoperatorFourier2}
\end{equation}

\begin{figure}
\begin{center}
\includegraphics[width=8cm,
keepaspectratio=true,clip=true]{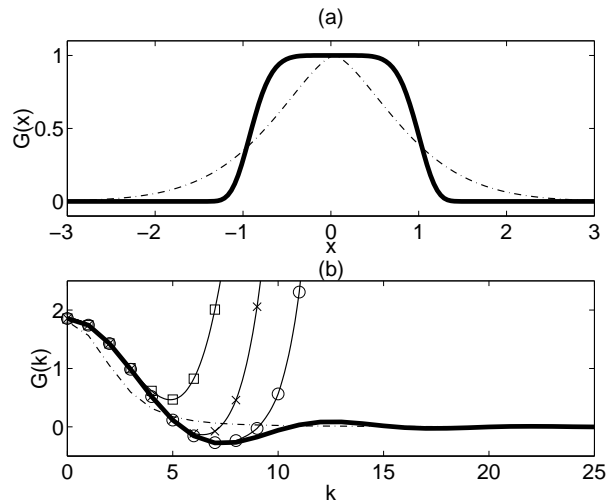}
\end{center}
\caption{Properties of the kernel (\ref{TypicalKernelShape}).
We choose $a=1$, $\sigma=1$ and $p$ significantly
smaller ($p=1.5$ for dash-dotted lines) or larger ($p=6$ for solid
lines) than $2$, to illustrate clearly the differences in
the Fourier transforms. (a) kernel functions $G(x)$ in niche space $x$,
(b) kernel functions in Fourier space $\hat{G}(k)$. Squares, crosses
and circles show the approximation (\ref{GoperatorFourier2}) to the $p=6$
kernel truncated after orders $k^4$, $k^8$, and $k^{12}$ respectively.
}
\label{KernelProperties}
\end{figure}

We illustrate the above manipulations with a relevant class of
competition kernels widely discussed in
\cite{Pigolotti2007,HernandezGarcia2009,Pigolotti2010}:
\begin{equation}
G(|x-s|) = a e^{-\left(\frac{|x-s|}{\sigma}\right)^p}, \\
\label{TypicalKernelShape}
\end{equation}
$p$ describes how steep the edge of the kernel is and $\sigma$
is the range of the competition. Note that $p=2$ is the
Gaussian kernel, and $p=1$ the exponential one. Species
consuming very different resources, i.e. with a large distance
between them in niche space ($|x-s|>>\sigma$) interact very
weakly, while species which are close ($|x-s|<\sigma$) compete
significantly. Finally, $a$ accounts for the strength of the
competition and sets the scale of the carrying capacity. In
Figure \ref{KernelProperties} we plot examples of the typical
kernel (\ref{TypicalKernelShape}) for two different values of
parameters, showing the important role of parameter $p$. The
Fourier transform of the function (\ref{TypicalKernelShape}) is
positive and tends to zero monotonously for $k \rightarrow
\infty$ when $p<2$ (see dash-dotted line), however for $p>2$
negative components appear in the Fourier transform (see solid
line) \cite{Bochner1937}. This leads to a modulational
instability of the homogeneous solution \cite{Pigolotti2007},
as detailed later. Figure \ref{KernelProperties} also shows,
for $p=6$, how the Taylor decomposition approaches the full
convolution kernel. The line marked by squares shows function
(\ref{GoperatorFourier2}) for a series of only three terms
($G_0, G_2$, and $G_4$). The line marked by crosses shows the
approximation with two more terms ($G_6, G_8$), and circles
show the approximation by terms up to $G_{12}$. The major
differences between Fourier transform of the full nonlocal
operator $\hat{G}(k)$ and the approximation
(\ref{GoperatorFourier2}) occur at high values of $k$. We keep
however only three terms in the series and perform the analysis
as an intermediate step towards understanding of the original
model. In this approximation the operator $\tilde{G}$ becomes
the Swift-Hohenberg operator or {\it shifted diffusion}, and
Eq. (\ref{KolmogorovFisherModel}) reduces to:
\begin{equation}
\label{SwiftHohenbergApproach}
\frac{\partial \Psi}{\partial t} =\Psi \left ( r - G_0 \Psi - G_2 \frac{\partial^2 \Psi}{\partial x^2} -
G_4 \frac{\partial^4 \Psi}{\partial x^4} \right )  + D \frac{\partial^2 \Psi}{\partial x^2},
\end{equation}
At difference with the original Swift-Hohenberg equation,
however, in this model the spatial operator appears in
nonlinear terms.

Within this approximation we interpret coefficients $G_0$,
$G_2$, and $G_4$, characterizing the kernel, as independent
parameters. This allows us to analyze the solutions of this
model more accurately and extract later the features a kernel
must have to access a given region of this parameter space.

\begin{figure}
\begin{center}
\includegraphics[width=8cm,
keepaspectratio=true,clip=true]{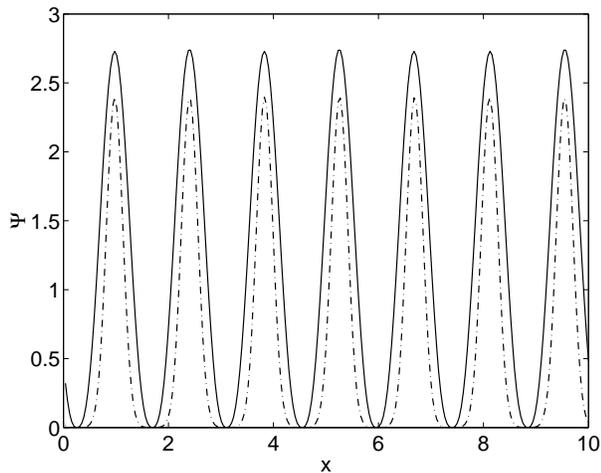}
\end{center}
\caption{Stationary density patterns $\Psi(x)$. Growth rate $r=1$,
diffusion coefficient $D=0.001$. Dash-dotted line is calculated using
the integral model (\ref{KolmogorovFisherModel}). The corresponding kernel
(\ref{TypicalKernelShape}) with $a=\sigma=1$ and $p=6$ is shown by the solid line in figure
\ref{KernelProperties}. The pattern plotted with a
solid line is calculated using the differential model (\ref{SwiftHohenbergApproach}),
with $G_0 = 2.15$, $G_2=0.36$, $G_4 =0.007$.}
\label{FigJustific}
\end{figure}

Truncating the nonlocal operator to obtain a local model can be
a rough approximation, however, it describes appropriately
stationary distributions $\Psi(x)$ provided $G_n k^n \hat\Psi(k)$
tends to zero fast enough as $k$ tends to infinity. Actually,
we find that a set of parameters $G_0, G_2, G_4$ close to the
ones obtained from (\ref{Gcoefficients}) gives a good
qualitative agreement between the stationary solutions of the
two models, as presented in Figure \ref{FigJustific}. The
patterns are calculated by solving Eqs.
(\ref{KolmogorovFisherModel}) and
(\ref{SwiftHohenbergApproach}), starting from slightly
(randomly) perturbed unstable homogeneous solutions as initial
conditions. In both cases we observe the formation of
``lumps'', separated by less populated regions. The similarity
of the results justifies the consideration of
(\ref{SwiftHohenbergApproach}) in the following sections. One
can note also that function (\ref{TypicalKernelShape}) decays
very fast for $|x-s| \rightarrow \infty$. This means that the
more narrow is the kernel the more local is the system, and the
validity of the truncation of the Taylor series is better. A quantitative 
evaluation of the effect of the truncation at a certain order can be obtained 
for each $k$ by comparing Eq. (\ref{GoperatorFourier}) with Eq. 
(\ref{GoperatorDecomposedFourier}).

Equations (\ref{KolmogorovFisherModel}) and
(\ref{SwiftHohenbergApproach}) have two homogeneous solutions:
$\Psi = 0$ and $\Psi = \Psi_0 = r/G_0$. The zero solution
corresponds to the situation in which niche space is not
occupied, and it is always unstable for positive growth rates
$r$. Any small number of individuals will be able to reproduce
and the population will grow approaching the steady homogeneous
state $\Psi_0$. In the absence of diffusion ($D=0$), this
solution is modulationally unstable for kernels whose Fourier
transform contains negative components
\cite{Lopez2004,Fuentes2004,Pigolotti2007}, and lumped
distributions over niche space arise instead. For kernels given
by Eq. (\ref{TypicalKernelShape}) negative Fourier components
appear when $p > 2$ \cite{Bochner1937}.

Adding diffusion $D > 0$, the stability condition is changed
and a threshold value appears. In the case of Eq.
(\ref{SwiftHohenbergApproach}), $\Psi_0$ is stable for
$G_2<G_2^{th}$, with

\begin{equation}
G_2^{th} = \frac{D G_0}{r} + 2 \sqrt{G_0 G_4}.
\label{ThresholdCondition}
\end{equation}

For $G_2>G_2^{th}$ a pattern with periodicity determined by the
critical wavenumber
\begin{equation}
k_c =\sqrt{\frac{G_0}{G_4}}
\end{equation}
appears.
For $D = 0$, condition (\ref{ThresholdCondition}) is equivalent to the
condition of appearance of negative components in the Fourier
transform of the kernel.

Since localized solutions are usually found in parameter
regions where a periodic pattern coexists with the homogeneous
solution \cite{Woods1999,Coullet2000}, we look for the
conditions in which the pattern-forming bifurcation is
subcritical. The way to do it (a weakly nonlinear analysis) is
described in \cite{Becherer2009,Manneville1990}. Introducing
formally a small parameter $\varepsilon$, we write the solution
and control parameter (we choose here $G_2$) as:
\begin{equation}
\Psi = \Psi_0 + \varepsilon \Psi_1 + \varepsilon^2 \Psi_2 + \varepsilon^3 \Psi_3 + ...
\label{PsiExp}
\end{equation}
\begin{equation}
G_2 = G_2^{th} + \varepsilon G_{21} + \varepsilon^2 G_{22} + \varepsilon^3 G_{23} + ...,
\label{G2Exp}
\end{equation}
where $\Psi_0$ is the homogeneous steady state. Substituting
(\ref{PsiExp}) and (\ref{G2Exp}) into the stationary version of
(\ref{SwiftHohenbergApproach}) and collecting terms at
different orders of $\varepsilon$ we obtain:
\begin{equation}
\left \{
\begin{array}{ll}
\varepsilon^0: & \Psi_0 (r  - G_{0} \Psi_0) = 0, \\
\varepsilon^1: & \tilde{L}_c \Psi_1 = 0,\\
\varepsilon^2: & \tilde{L}_c \Psi_2 = f_2,\\
\varepsilon^3: & \tilde{L}_c \Psi_3 = f_3,
\end{array}
\right .
\label{EpsilonOrders}
\end{equation}
where
\begin{equation}
\tilde{L}_c = \Psi_0 \left( -G_0 - G_2^{th}\frac{\partial^2}{\partial x^2} - G_4 \frac{\partial^4}{\partial x^4} \right) + D \frac{\partial^2}{\partial x^2}
\end{equation}
is the Jacobian of Eq. (\ref{SwiftHohenbergApproach}) evaluated at $G_2=G_2^{th}$, and
\begin{equation}
f_2 = \frac{D}{\Psi_0} \Psi_1 \frac{\partial^2}{\partial x^2} \Psi_1 + \Psi_0 G_{21} \frac{\partial^2}{\partial x^2}\Psi_1,
\label{f2}
\end{equation}
\begin{equation}
\begin{array}{l}
f_3 = \Psi_0 G_{22} \frac{\partial^2}{\partial x^2} \Psi_1 + \frac{D}{\Psi_0} \Psi_2 \frac{\partial^2}{\partial x^2} \Psi_1 +  \\
\qquad \qquad + \frac{D}{\Psi_0} \Psi_1 \frac{\partial^2}{\partial x^2} \Psi_2 - \frac{D}{\Psi_0^2}\Psi_1^2 \frac{\partial^2}{\partial x^2}\Psi_1 - \\
\qquad \qquad \qquad \qquad + G_{21} \Psi_0 \frac{\partial^2}{\partial x^2} \Psi_2.
\end{array}
\label{f3}
\end{equation}

At first order $\Psi_1 = A \cos(k_c x)$, which is the periodic
solution bifurcating at $G_2=G_2^{th}$. At second order, the
solvability condition
\begin{equation}
\int \limits_{0}^{2\pi/k_c} f_2 \Psi_1 dx = 0,
\label{orthogonality1}
\end{equation}
leads to $G_{21}=0$ and $\Psi_2 = B + C \cos(2 k_c x)$, with
$B= D G_0 \sqrt{G_0} A^2/2 r^2 \sqrt{G_4}$, and $ C = D G_0 \sqrt{G_0} A^2/18 r^2 \sqrt{G_4} $.

Finally, the solvability condition at third order
\begin{equation}
\int \limits_{0}^{2\pi/k_c} f_3 \Psi_1 dx = 0
\label{orthogonality2}
\end{equation}
leads to the following equation for the stationary amplitude
$A$ of the critical mode found at the first order:
\begin{equation}
- \frac{2 G_{22} r} {G_0} A + \kappa A^3 = 0,
\label{ampleqA}
\end{equation}
where
\begin{equation}
\kappa=\frac{3 D G_0^2}{2 r^2} - \frac{23 D^2 G_0^2 \sqrt{G_0}} {18 r^3 \sqrt{G_4}}.
\label{kappa}
\end{equation}
The transition from a super to a sub-critical bifurcation
occurs when the coefficient $\kappa$ changes sign ($\kappa=0$).
This happens for
\begin{equation}
D = D_s =\frac{27}{23} \frac{r \sqrt{G_4}}{\sqrt{G_0}}.
\label{SubcriticalityMain}
\end{equation}
For the full nonlocal operator, this condition is equivalent to
setting the coefficient $\kappa$ of Eq. (28) in
Ref.~\cite{Lopez2004} to zero.

\section{Self-localized solutions}
\label{results}

In the following we study the existence of localized solutions
in Eq. (\ref{SwiftHohenbergApproach}). This equation has only
two independent parameters, so that by rescaling $t$, $x$, and
$\Psi$ we can consider $G_0=1$, $G_4=1$, and $r=1$ without loss
of generality, and take $G_2$ and $D$ as control parameters.
The condition for instability of the homogeneous solution
(\ref{ThresholdCondition}) becomes then:
\begin{equation}
G_2 > G_2^{th} = D + 2,
\label{instabthresh}
\end{equation}
and the pattern appears subcritically if
\begin{equation}
D > D_s = \frac{27}{23}.
\label{SubcriticalityCondition}
\end{equation}

\begin{figure}
\includegraphics[width=8cm,
keepaspectratio=true,clip=true]{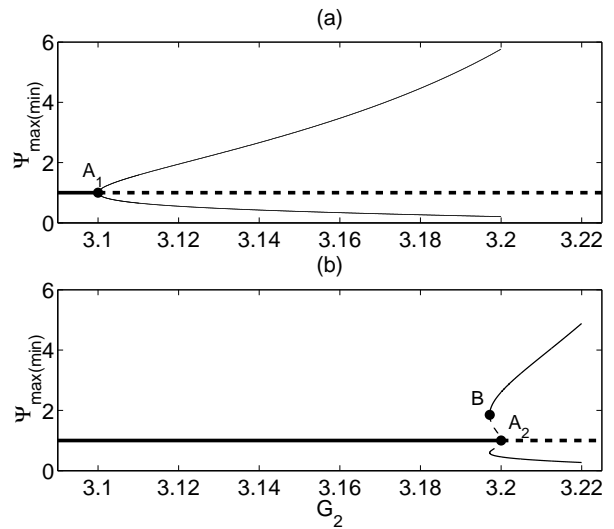}
\caption{Bifurcation diagram of the stationary periodic-pattern solution (see Fig.~\ref{FigJustific}) of
Eq. (\ref{SwiftHohenbergApproach}) for two values of
$D$, (a) $D=1.1$ i.e. below $D_s$  and (b) $D=1.2$ i.e. above $D_s$. The thin solid (dashed) lines show the maximum and
minimum values of the stable (unstable) solutions.
The bold solid (dashed) line represents the stable (unstable) homogeneous solution, which does not
depend on $G_2$. Points $A_1$ and $A_2$ indicate the instability thresholds,
given by (\ref{instabthresh}). $B$ indicates the turning point of
the subcritical bifurcation. $r=G_0=G_4=1$.}
\label{SubSuperFigure}
\end{figure}

To illustrate the change from a supercritical to a subcritical
bifurcation we plot the bifurcation diagram of the stationary
pattern solution of (\ref{SwiftHohenbergApproach}) arising at
$G_2^{th}$ for two values of $D$, one below and one above the
critical value $D_s$ (See Fig. \ref{SubSuperFigure}).

The codimension-2 point indicated by $D=D_s$ and $G_2=G_2^{th}$
is called in the spatial dynamics parlance a Degenerate
Hamiltonian-Hopf bifurcation, and it is known to be the origin
of the existence of localized states in pattern forming systems
\cite{Woods1999}. In the following we focus on the existence of
self-localized states consisting on a number of stable lumps of
the pattern solution on top of the homogeneous solution
$\Psi_0$. For this we move well into the parameter region where
the pattern forming bifurcation is subcritical by increasing
$D$.

\begin{figure}
\includegraphics[width=8cm,
keepaspectratio=true,clip=true]{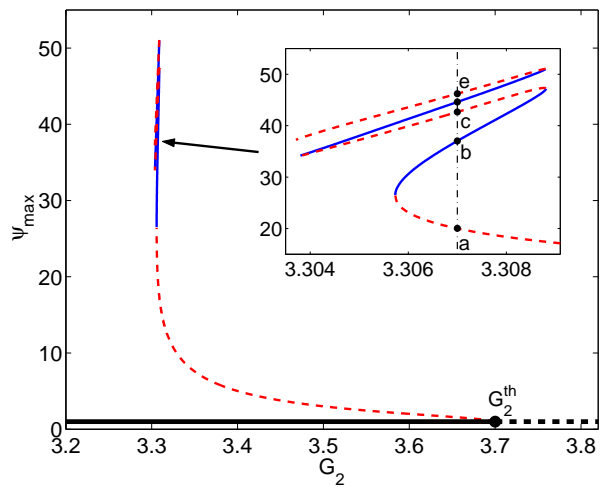}
\caption{(Color online) Branch of localized states of model 
(\ref{SwiftHohenbergApproach}) for $D=1.7$.
The maximum of $\Psi(x)$ as a function of $G_2$ is shown.
Solid (dashed) lines indicate stable (unstable solutions).
The inset shows a zoom of the region of existence of localized
states displaying the typical snaking. The vertical dash-dotted
line indicates $G_2=3.307$, value for which examples of such solutions
(a-e) are shown in Fig. \ref{SolitonExamplesFigure}. For clarity in the plot we 
do not display the label d corresponding to the point between c and e.
Other parameters: $r=G_0=G_4=1$.
} \label{FigSolBif}
\end{figure}

\begin{figure}
\includegraphics[width=8cm,
keepaspectratio=true,clip=true]{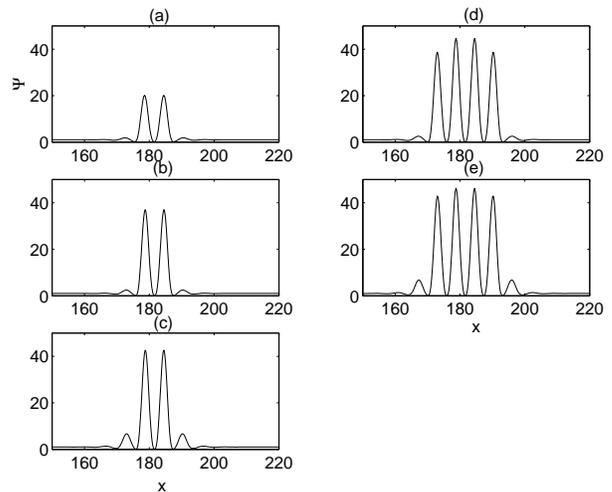}
\caption{Examples of localized states sitting on nonzero homogeneous background
of model (\ref{SwiftHohenbergApproach}) for parameters corresponding to
the marked points in the inset of Figure \ref{FigSolBif}.
The state (a) is the separatrix between attraction basins of the
homogeneous solution and the state (b). The state (c) is accordingly
between the solutions (b) and (d).}
\label{SolitonExamplesFigure}
\end{figure}

Using a shooting method (see Appendix) where the spatial
coordinate $x$ is used as a dynamical variable in the
stationary version of Eq. (\ref{SwiftHohenbergApproach}), we
have found a self-localized solution. Taking it as an initial
guess we have computed its branch by continuation techniques
using a Newton method. Figure \ref{FigSolBif} shows the
bifurcation diagram of localized states with an even number of
peaks for $D=1.7$. Another analogous curve for localized states
with odd number of peaks (not shown) also exists. The curve
shows a characteristic snaking structure. The branch follows a
series of saddle-node bifurcations that transform unstable
solutions into stable localized states, adding each time a peak
at each side of the structure as one moves up. Typical
localized states, indicated by bold dots in the inset of Figure
\ref{FigSolBif}, are presented in Figure
\ref{SolitonExamplesFigure}. Solutions (b) and (d) in figure
\ref{SolitonExamplesFigure} are stable, while the rest are
unstable. As can be seen from the inset in Figure
\ref{FigSolBif}, the region of existence of stable
self-localized states is very narrow due to the proximity to
the codimension-2 point. This region becomes larger as one
moves away from this point in the direction of increasing the
subcriticality of the pattern, i.e. increasing $D$. However we
can not go much further into that region, because the minimum
of the population distribution approaches the trivial zero
solution too much, and our simulations diverge. In this case
further analysis is not possible and saturating terms should
probably be included in Eq. (\ref{KolmogorovFisherModel}) in
order to observe stable localized states in wider parameter
regions. We are unable to determine if the difficulties are
only of numerical origin or if there is some more fundamental
change of behavior or bifurcation when increasing
subcriticality, perhaps associated to some spatial analog of
the {\sl paradox of enrichment} \cite{Rosenzweig1971}. Further
investigation is needed to clarify this point.

\begin{figure}
\includegraphics[width=8cm,
keepaspectratio=true,clip=true]{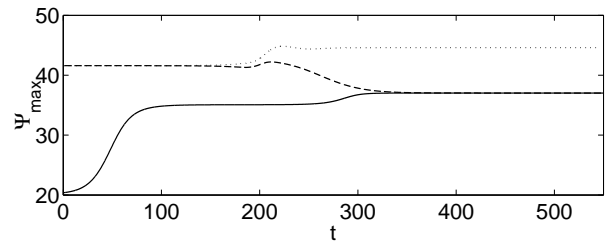}
\caption{Temporal dynamics of the maximum of $\Psi$. The dotted (dashed) line
show the transition to state (d)[(b)] in Fig. \ref{SolitonExamplesFigure}
starting from a state slightly above (below) (c). The solid line shows the
transition to (b) starting from a state slightly above (a).}
\label{FigDyn}
\end{figure}

The localized solutions found (Fig. \ref{SolitonExamplesFigure})
consist on a few lumps of species, of a very high population
density, which locally deplete close niche positions but do not
make them completely empty. Further apart the effect of the
lumps becomes unimportant and the density in the rest of the
system consists on the stable homogeneous coexistence of
species given by the homogeneous solution $\Psi=1$. The spacing
between the lumps forming the localized patch is of the order
of the periodicity of the extended pattern (Fig.
\ref{FigJustific}). These regions can be then considered as
portions of the periodic pattern embedded inside the stable
homogeneous solution. To illustrate the stability of the self-localized
states (b) and (d) in Fig. \ref{SolitonExamplesFigure}, we show in Figure
\ref{FigDyn} the switching dynamics of localized states starting from suitable
initial conditions. The stability of the states has also been checked with
respect to small additive noise.

\section{Localized states in the full nonlocal model}
\label{backtononlocal}

Once the precise conditions for the observation of subcritical
patterns and localized structures have been determined for Eq.
(\ref{SwiftHohenbergApproach}), we can discuss the implications
for the kernel in the full nonlocal model
(\ref{KolmogorovFisherModel}). The main assumption is that we
can apply to (\ref{KolmogorovFisherModel}) the results of the
previous section by using the values $G_0$, $G_2$, etc. arising
from the expansion of the nonlocal kernel.

The necessary conditions for the existence of self-localized
states were (for $G_0=G_4=r=1$) the subcriticality
criterion (\ref{SubcriticalityCondition}), $D > D_s = 27/23$,
and  the instability condition (\ref{instabthresh}),
$G_2 > G_2^{th} = D + 2$. In addition, the subcritical region
increases with increasing $D$, but as commented above our
numerical results were unable to probe large values of $D$
without divergences. Figure (\ref{FigSolBif}) illustrate the
situation for one of the largest values of $D$ attainable,
$D=1.7$, for which localized solutions appear for
$G_2\approx [3.30,3.31]$.

But it happens that this range of values of $G_2$ is far from
what is achievable with an interaction kernel of competitive
nature exclusively (i.e. one taking only positive values). To
see this we note that a positively-defined and normalized
($G_0=1$) kernel can be interpreted as a probability density,
so that $G_2$ and $G_4$ are its moments (see
Eq.~(\ref{Gcoefficients})): $G_2=\langle x^2 \rangle /2$,
$G_4=\langle x^4 \rangle/24$. If $G_4=1$, then $\langle x^4
\rangle=24$. Applying the moment monotonicity inequality:
$\langle |x|^r \rangle^{1/r} \le \langle |x|^s \rangle^{1/s}$,
where $0 < r \le s$, and using $r=2$ and $s=4$, we have
$(2G_2)^{1/2} \le 24^{1/4}$, or $G_2 \le \sqrt{6}\approx
2.449$. This limiting value is well below the ones needed to
observe self-localized solutions of Eq.
(\ref{SwiftHohenbergApproach}) without encountering
divergences. We can not completely discard in a rigorous manner
the possibility of stable localized structures to exist for the
nonlocal model at sufficiently large values of $D$, nor the
presence or other localized solution branches not captured
within the differential approximation. But the fact is that we
have been unable to find numerically self-localized solutions
of (\ref{KolmogorovFisherModel}) when using a purely
competitive (i.e. positive) kernel $G(x)$.

A natural way to achieve the larger values of $G_2$ needed is
to allow the kernel to take negative values close to $x=0$ or
for large values of $x$. This means the presence of
facilitative interactions (mutualism, symbiosis, ...) together
with the competitive ones. We note that such combination of
positive and negative interactions at different length scales
was already proposed from biological reasoning in an early
paper \cite{Britton1989}, an it is an important ingredient in
the modeling of vegetation patterns \cite{Borgogno2009}. Here
we consider an integral kernel $G_I$ of the form:

\begin{equation}
G_I(|x-s|) = a_1 e^{-\left(\frac{|x-s|}{\sigma_1}\right)^{p_1}} + a_2
e^{-\left(\frac{|x-s|}{\sigma_2}\right)^{p_2}}, \\
\label{KernelI}
\end{equation}
were $a_1$ can take negative values modelling cooperative or
facilitative interactions.

To find localized states in the full nonlocal model we perform
then a continuation of the localized states from the
differential to the integral kernel. The main difference
between these two cases consists in the behavior of $G(k)$ for
$k \rightarrow \infty$: in the differential case,
$G(k)\rightarrow \infty$, while for the integral case
$G(k)\rightarrow 0$, as illustrated in Fig.
(\ref{KernelProperties}). Since the stability range of
localized states is so small it is a challenge to find the
parameters of the nonlocal kernel that support stable localized
states. We show now, however, that at least for the lowest
unstable localized state marked by a dashed red line in Figure
(\ref{FigSolBif}) our continuation strategy is able to find
them. To do so, we consider a linear combination of the
integral kernel $G_I$ and the differential approximation in the
truncated model (\ref{SwiftHohenbergApproach}) $G_D$ in the
form:

\begin{equation}
\label{FromDiffToIntegro}
G(k) = \gamma G_D(k) + (1-\gamma) G_I(k),
\end{equation}
where $\gamma$ is a parameter characterizing how differential
or how integral the resulting kernel $G(k)$ is. So, for
$\gamma=1$, the kernel is purely differential, while for
$\gamma=0$, the kernel is purely integral.

\begin{figure}
\includegraphics[width=8cm,
keepaspectratio=true,clip=true]{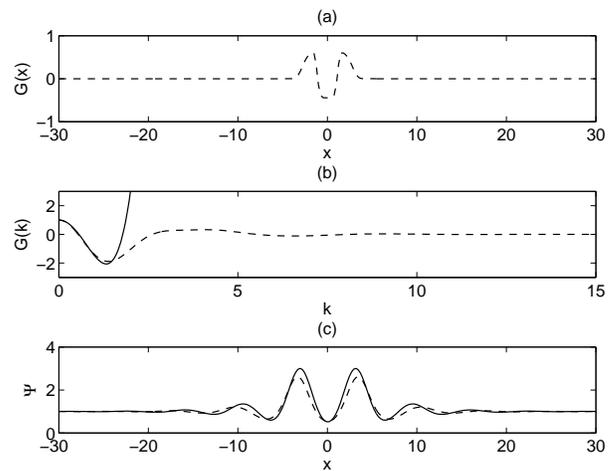}
\caption{Transition from the differential to the integral kernel.
(a) Kernel $G_I(x)$ with $a_1 = -1.07561$, $p_1 = 6.0$, $\sigma_1 =
1.2$, $a_2 = 0.63103$, $p_2 = 6.0$, and $\sigma_2 = 2.9$.
(b) Dashed line - kernel $G_I(k)$. Solid line - differential 
approximation $G_D(k)$ with $G_0=1.0$, $G_2=3.5$, $G_4=1.0$.
(c) Profiles of localized solutions for the respective kernels of panel (b).
Other parameters are: $D = 1.7$, $r=1.0$.}
\label{FigKernels1}
\end{figure}

We choose the parameters in such a way that the Fourier
transform of $G_I$ is very close to the one of $G_D$ used in
Fig. \ref{FigSolBif}, except for high values of $k$ [see Figure
\ref{FigKernels1}(b)]. This implies that in real space the
kernel $G(x)$ takes negative values close to $x=0$ [
Figure \ref{FigKernels1}(a)] . Now, using
a continuation method we follow the self-localized solution
from the differential case $\gamma=1$ to the integral case
$\gamma=0$. The corresponding modification of the kernel
(\ref{FromDiffToIntegro}) and of the profile of the localized
separatrix solution is shown in Fig. \ref{FigKernels1}(c). In
such a way we demonstrate the existence of localized states in
the original model (\ref{KolmogorovFisherModel}) for kernels
fulfilling appropriate conditions.

\section{Conclusions}
\label{conclusions}

By studying a differential truncation of the nonlocal
Kolmogorov-Fisher model we have rigorously calculated the
conditions by which periodic patterns are subcritical and we
have demonstrated numerically the existence of the stable
localized states in the differential approximation. These are
patches of finite extent containing a number of lumps of
species and arise on top of the homogeneous distribution. In
contrast to other works, our results show that the necessary
ingredient to observe stable self-localized states, namely the
presence of subcritical patterns, is already present in systems
with spatial coupling in the quadratic nonlinearity only,
rather than nonlinearities of different orders being necessary.
In consequence, the localized patches appear on a background of
homogeneously distributed species coexistence, instead than on
top of the no-species empty state. Extending the results
obtained for the truncation to the full nonlocal model we find,
however, that competitive interactions alone can not lead to
the conditions for the observation of localized states, and
facilitative interactions at $x=0$ or with distant locations in
niche space, modeled by negative values of the kernel, are
needed to observe this phenomenon.

From a biological point of view, the self-localization
indicates that species with no particular advantage may
predominate to competitors. A patch of species can be formed at
any position of niche space by a particular initial condition
or temporary perturbation. One should note that inhomogeneities
in $r$ could increase or decrease the stability of the
considered states. Although the results have been obtained in
one dimensional space, and there are important differences with
higher dimensional cases, we expect that the conditions for the
observation of localized states will be qualitatively similar.

We acknowledge financial support from FEDER and MINECO (Spain)
through grant
FIS2012-30634 INTENSE@COSYP, and from Comunitat Aut\'onoma
de les Illes Balears. DG acknowledges support from CSIC  (Spain) through grant
number 201050I016.

\section*{Appendix: Numerical methods} \label{AppendixA}

To find the self-localized solution of (\ref{SwiftHohenbergApproach})
we write first the steady state condition $\frac{\partial}{\partial t} =0$:
\begin{equation}
\label{StationaryEq}
0 =\Psi \left ( r - G_0 \Psi - G_2 \frac{\partial^2 \Psi}{\partial x^2} -
G_4 \frac{\partial^4 \Psi}{\partial x^4} \right )  + D \frac{\partial^2 \Psi}{\partial x^2},
\end{equation}
introducing auxiliary quantities $a,b,c$
equation (\ref{StationaryEq}) is transformed to the system of ordinary
differential equations:

\begin{equation}
\begin{array}{l}
\frac{\partial \Psi}{\partial x} = a, \\
\\
\frac{\partial a}{\partial x} = b, \\
\\
\frac{\partial b}{\partial x} = c, \\
\\
\frac{\partial c}{\partial x} = -\frac{1}{G_4} \left ( r - G_0 \Psi + G_2 b + \frac{D b}{\Psi} \right ).
\end{array}
\label{ODEapproach}
\end{equation}

\begin{figure}
\begin{center} 
\includegraphics[width=8cm,keepaspectratio=true,clip=true]{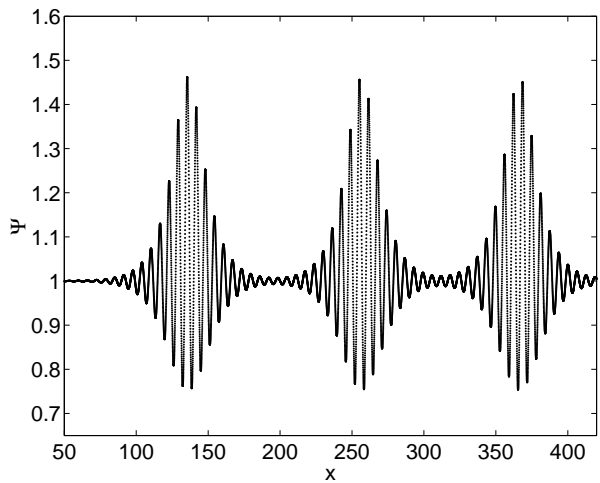} 
\end{center}
\caption{Dynamics of the model (\ref{ODEapproach})
with parameters of Figure. \ref{FigSolBif} and $G_2=3.67$.}
\label{SepSolFigure}
\end{figure}

Interpreting now $x$ as a dynamical variable we solve the system
(\ref{ODEapproach}) with initial conditions $a=0$, $b=0$, $c=0$, and
$\Psi=\Psi_0=1$ plus small
perturbations. For parameters close to the subcritical bifurcation, trajectories
showing localized pulses as the ones shown in Figure
\ref{SepSolFigure} are easily found. Taking one of the chirped
pulses of this figure as a initial guess we can compute the branch shown in
Fig. \ref{FigSolBif} using a Newton method and continuation
techniques \cite{NumRec}.

\end{document}